\documentclass[onecolumn,showpacs,aps,epsfig,nofootinbib]{revtex4}

\usepackage{graphicx}
\usepackage{epstopdf}
\usepackage{latexsym}

\def\be{\begin{equation}}
\def\ee{\end{equation}}
\def\bea{\begin{eqnarray}}
\def\eea{\end{eqnarray}}

\newcommand{\la}{\lambda}

\newcommand{\bear}{\begin{eqnarray}}

\newcommand{\eear}{\end{eqnarray}}
\newcommand{\lrp}[1]{\left( #1 \right)}  

\newbox\pippobox

\def\be{\begin{equation}}
\def\ee{\end{equation}}
\def\bea{\begin{eqnarray}}
\def\eea{\end{eqnarray}}
\def\bx{{\bf x}}
\def\bk{{\bf k}}

\def\ve{{\bf e}}
\def\a{\alpha}
\def\e{\epsilon}
\def\m{\mu}

\def\r{\rho}
\def\s{\sigma}

\def\z{\zeta}
\def\b{\beta}

\def\h{\eta}
\def\d{\delta}
\def\p{\phi}

\def\nn{\nonumber}
\def\half{\frac12}
\def\le{\left}
\def\ri{\right}
\def\6{\partial}

\def\vp{\varphi}
\def\f{\frac}
\def\ma{\mathcal}
\def\la{\langle}
\def\ra{\rangle}
\def\B{\Big}
\def\Mp{M_{pl}}
\def\tld{\tilde}

\begin{document}

\title{{\bf Entropy Perturbations in N-flation}}

\author{Rong-Gen Cai$^{1,}$\footnote{Email: cairg@itp.ac.cn}}

\author{Bin Hu$^{1,}$\footnote{Email: hubin@itp.ac.cn}}

\author{Yun-Song Piao$^{2,}$\footnote{Email: yspiao@gucas.ac.cn}}

\affiliation{$^{1}$Key Laboratory of Frontiers in Theoretical
Physics, Institute of Theoretical Physics, Chinese Academy of
Sciences, P.O. Box 2735, Beijing 100190, China\\
$^{2}$College of Physical Sciences, Graduate School of Chinese
Academy of Sciences, Beijing 100049, China}

\date{\today}

\begin{abstract}

In this paper we study the entropy perturbations in N-flation by
using the $\d\ma{N}$ formalism. We calculate the entropy
corrections to the power spectrum of the overall curvature
perturbation $P_{\z}$.  We obtain an analytic form of the transfer
coefficient $T^2_{\ma{R}\ma{S}}$, which describes the correlation
between the curvature and entropy perturbations, and investigate
its behavior numerically. It turns out that the entropy
perturbations cannot be neglected in N-flation, because the
amplitude of entropy components is approximately in the same order
as the adiabatic one at the end of inflation
$T^2_{\ma{R}\ma{S}}\sim\ma{O}(1)$. The spectral index $n_S$ is
calculated and it becomes smaller after the entropy modes are
taken into account, i.e., the spectrum becomes redder, compared to
the pure adiabatic case. Finally we study the modified consistency
relation of N-flation, and find that the tensor-to-scalar ratio
($r\simeq0.006$) is greatly suppressed by the entropy modes,
compared to the pure adiabatic one ($r\simeq0.017$) at the end of
inflation.

\end{abstract}


\maketitle


\section{Introduction}

Inflation is now a standard paradigm for describing the physics of
the very early universe, but the microphysics nature of the
field(s) responsible for inflation remains unknown. In the last
few decades intensive effort has been devoted to understanding the
fundamental physics of the inflation theory. For simplicity, most
of studies have been focused on the effective single scalar field
model, however, in the low-energy limit of string theory, more
than one scalar fields are present and they may work cooperatively
to drive the inflation, such as the assisted inflation
\cite{assisted}.

Recently, Dimopoulos {\it et~al.} \cite{Dimopoulos:2005ac} showed
that the many axion fields predicted by string vacua can be
combined and lead to a radiatively stable inflation, called
N-flation. Using the random matrix theory Easther and McAllister
\cite{Easther:2005zr} showed that the mass distribution for $N$
axion fields should be in the Marc\v enko-Pastur spectrum form.
Further, many cosmological observable imprints of N-flation have
been investigated, such as the tensor-to-scalar ratio $r$
\cite{Alabidi:2005qi}, the non-Gaussianity parameter $f_{{\rm
NL}}$ \cite{NflationNG,Battefeld:2006sz}, and the scalar spectral
index $n_S$ \cite{ns} for the pure adiabatic perturbation. The
results show that for $r$ and $f_{{\rm NL}}$ the deviations from
the single-field models are negligible, however, the spectral
index $n_S$ is smaller  than the case of the single-field models.
The preheating process after N-flation is numerically investigated
in \cite{Battefeld:2008bu} and the results show that the
parametric resonance is suppressed which differs significantly
from the single-field case.

Compared with the single field model, the presence of multiple
fields during inflation can lead to quite different inflationary
dynamics \cite{Christopherson:2008ry} (see \cite{Wands:2007bd} for
a review). In particular, multiple fields can lead to the
generation of entropy (non-adiabatic) perturbations during
inflation, which can alter the evolution of the overall curvature
perturbation \cite{Gordon:2000hv} and produce a detectable
non-Gaussianity \cite{Gao:2009gd}. For a two-field model the
entropy perturbations are investigated both analytically and
numerically \cite{twofield}, however, the generalization from the
two-field model to the model with a large number of fields is less
developed. For the N-flation model the entropy perturbations have
been investigated by different approaches
\cite{Battefeld:2006sz,Choi:2008et}. By virtue of the $\d\ma{N}$
formalism \cite{dN}, an analytic form of spectral index is derived
in \cite{Battefeld:2006sz}, and similar result is obtained by the
authors of \cite{Choi:2008et} by using a different approach
\cite{Polarski:1994rz}. In this paper, using the new
interpretations of the $\d\ma{N}$ formalism which are developed in
\cite{Tye:2008ef}, we calculate the entropy corrections to the
power spectrum of the overall curvature perturbation,
corresponding spectral index and the tensor-to-scalar ratio. Our
numerical results are in agreement with the earlier results in
\cite{Battefeld:2006sz,Choi:2008et}.

This paper is organized as follows. In Sec.~\ref{nflation} we
briefly review the constructions of N-flation and the mass
distribution of $N$ axion fields, then investigate the
cosmological background evolutions numerically. In
Sec.~\ref{linear} we study the linear perturbation, derive
explicitly the entropy corrections to the primordial power
spectrum, and then calculate the power spectrum, spectral index
and the tensor-to-scalar ratio numerically for the N-flation
model. Sec.~\ref{concl} is devoted to our conclusions.

\section{\label{nflation} Review of N-flation}

In this section we  briefly review the construction of N-flation,
especially focus on the quadratic potential and the mass spectrum.
Then we investigate the cosmological background dynamics with a
given mass distribution.

\subsection{Quadratic potential in N-flation}

Dimpopoulos ${\it et~al.}$ in~\cite{Dimopoulos:2005ac} consider a
potential of $N$ axions as
 \be
 \label{periodic potential}
 V(\vp_1,\vp_2,\cdots,\vp_N)=\sum_{I=1}^NV_I(\vp_I)=\sum_{I=1}^N
 \Lambda^4_I\Big(1-\cos\Big(\f{\vp_I}{f_I}\Big)\Big)\;.\ee
where $V_I$ is the periodic potential which arises solely from
non-perturbative effects, $f_I$ is the axion decay constant and
$\Lambda_I$ is the dynamically generated scale of the axion
potential that typically arises from an instanton expansion. This
scale can be many orders of magnitude smaller than the Planck scale.

For small field values $\vp_I\ll M_{pl}$ the periodic potentials can
be Taylor expanded as
 \be
 \label{taylor exp}
 V_I(\vp_I)\simeq\half m_I^2\vp_I^2+\cdots\;,\ee
with the masses $m_I^2=\Lambda_I^4/f_I^2$. Consider the case in
which the masses $\{m_I\}$ are distributed uniformly and the axion
fields start out displaced from the minimum by
$\vp_{I0}=\a_IM_{pl}$, with the maximum displacement set by each
axion decay constant
 \be
 \a_I^2\leq\f{f_I^2}{M_{pl}^2}\;,\ee
then it is effectively equivalent to the scenario of a single
field $\Phi$ with a super-Planckian displacement $\sqrt{N}\a
M_{pl}$. This means that the typical initial condition in the
large $N$ limit is expected to be super-Planckian and it is
suitable for chaotic inflation. In this sense, N-flation realizes
the  $m^2\Phi^2$ inflation in a very well-controlled string theory
setting.

The authors of \cite{Dimopoulos:2005ac} assume a uniform axion
mass spectrum for simplicity, however, for a realistic model we
should exactly determine which sorts of mass spectra are possible
in string compactification. Surprisingly, in the $N\to\infty$
limit, using the random matrix theory Easther and McAllister
\cite{Easther:2005zr} obtained an essentially universal mass
spectrum, without invoking details of the compactification, such
as the intersection numbers, the choice of fluxes, or the location
in moduli space.

\subsection{Mass spectrum}

 Consider the Lagrangian
of axions with kinetic and potential terms as
 \be
 \label{lag1}
 \ma{L}=\half M_{pl}^2K_{IJ}\nabla_{\mu}\vp^I\nabla^{\mu}\vp^J-V\;,\ee
where the supergravity potential and the KKLT superpotential
reads~\cite{Kachru:2003aw}
 \bea
 \label{sg}
 V&=&\exp\Big(\f{K}{M_{pl}^2}\Big)\Big(K^{AB}D_AWD_{\bar B}\bar W-3\f{|W|^2}{M_{pl}^2}\Big)\;,\\
 \label{kklt}
 W&=&W_0(S,\chi_a)+\sum_IA_I(\chi_a)\exp\Big\{-a_I(\rho_I-i\vp_I)\Big\}\;,\eea
with $A,~B$ run over the dilaton $S$, the complex structure moduli
$\chi_a$, and the K\"ahler moduli $\rho_I$. Inserting (\ref{kklt})
into (\ref{sg}),  expanding the potential around the origin
$\vp_I=0$, and using the F-flatness conditions $D_AW|_{\vp_I=0}=0$,
we have
 \be
 V=(2\pi)^2\hat{M}_{IJ}\vp^I\vp^J+\cdots\;,\ee
where the mass matrix is~\footnote{Here we emphasize that the moduli
$\chi_a$, $\rho_I$, which appear in $C_I\equiv A_Ie^{-2\pi\rho_I}$,
are not dynamical variables.}
 \be
 \label{mass matrix}
 \hat M_{IJ}=\f{1}{M_{pl}^2}e^K\Big(K^{AB}D_AC_ID_BC_J-3C_IC_J\Big)\;.
 \ee
From (\ref{lag1}) and (\ref{mass matrix}), it is easy to see that
the kinetic terms and the mass matrix are obviously  not diagonal in
the basis where the superpotential is simple.  We now perform two
orthogonal rotations to diagonalize $K_{IJ}$ and $\hat M_{IJ}$.
First, we rotate the basis to make the axion kinetic terms be
canonical
 \be
 \label{lag2}
 \ma{L}=\half\6_{\m}\vp_I\6^{\m}\vp^I-M_{IJ}\vp^I\vp^J\;,\ee
where the mass matrix becomes
 \be\label{mass matrix2}
 M_{IJ}=(2\pi)^2\f{e^K}{f_If_J}O_I^M\B(D_AC_MD^AC_L-3C_MC_L\B)O_J^L\;.\ee
We  perform the second orthogonal rotation so that $M_{IJ}$ is
diagonalized, because the potential simply takes a purely quadratic
form. In order to reach a more clear result, it is helpful to
introduce a new $(N+P)\times N$ rectangular matrix
 \be\label{R_Ai}
 R_{AI}\equiv 2\pi e^{K/2}f_I^{-1}O_I^J(D_AW_J)\;,\ee
then the mass matrix $M_{IJ}$ becomes
 \be\label{mass matrix3}
 M_{IJ}=R_{IA}R_{\ J}^A\;.\ee
Because we do not know about the individual terms $D_AC_I$, in what
follows we will regard them as random variables, i.e., we
 take $M_{IJ}$ as a random matrix. Now the task is to
 determine the eigenvalue spectra of  the $N\times N$
random matrix. Surprisingly, with the random matrix theory, authors
of \cite{Easther:2005zr} find that in the large $N$ limit the mass
spectrum is independent of concrete values of $D_AC_I$ and then
obtain an essentially universal mass spectrum
 \be
 \label{mp}
 \ma P(m^2)=\f{1}{2\pi m^2\b\s^2}\sqrt{(b-m^2)(m^2-a)}\;,\ee
for $a\leq m^2\leq b$, where $\b=N/(N+P)$ is the ratio of the
dimensions of the rectangular matrix $R$, $\s^2$ is the variance of
the entries of $R_{AI}$ and $a$, $b$ are defined as
 \bea
 a&=&\s^2\Big(1-\sqrt{\b}\Big)^2\;,\\
 b&=&\s^2\Big(1+\sqrt{\b}\Big)^2\;.
 \eea
The spectrum (\ref{mp}) is nothing, but the Marc\v enko-Pastur
(MP) spectrum. The normalized MP spectrum (\ref{mp}) describes the
mass distribution probability of a single axion field. On the
other hand, the law of large numbers ensures that the mass
distribution of $N$ axion fields obeys the distribution
probability of the single field. In practice we uniformly split
the mass range of $N$ axions $(a,b)$  into $\tld N$ ($\tld N\ll
N$) bins
 \be\label{bins}(\tld m_0^2,\tld m_1^2),(\tld m_1^2,\tld m_2^2),\cdots,
 (\tld m_{\tld N-1}^2,\tld m_{\tld N}^2)\;,\ee
where $\tld m_I^2$ and the width of each bin $\d$ are
 \be\tld m_0^2=a\;,\qquad \tld m_{\tld N}^2=b\;,\qquad \tld m_I^2=\tld m_{I-1}^2+\d\;,\qquad
 \d=(\tld m_{\tld N}^2-\tld m_0^2)/\tld N\;.\ee
 \begin{figure}[h]
    \centering
    \includegraphics[angle=-90,width=9.5cm]{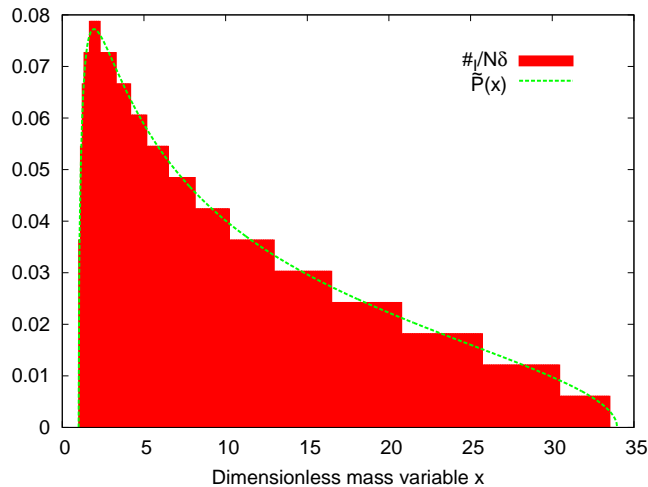}
    \caption{This figure illustrates the mass distribution of $N=1500$ axions
    versus the dimensionless mass variable $x_I\equiv m_I^2/a$, in the case of $\beta=1/2$.
            The red boxes denote the quantity $\#_I/N\d$ in the $I$th bin,
            and the green smoothed curve is the normalized mass distribution
            probability $\tld \ma P(x_I)$ for a single axion field (Marc\v enko-Pastur distribution).
            Note that the law of large numbers ensures that the mass distribution of $N$
            axions obeys the distribution probability of a single field.}
    \label{mass}
\end{figure}
Furthermore we set the masses of the axions in each bin as
 \be\label{binmass}m_I^2=(\tld m_{I-1}^2+\tld m_{I}^2)/2\;,\qquad I=1,2,\cdots,\tld N\;.\ee
Due to the law of large numbers we then have the following
relation
 \be\label{large number}\f{\#_I}{N}=\ma P(m_I^2)\d\;,\ee
where $\#_I$ denotes the number of axions in the $I$th bin.

Note that the constraint from the renormalization of Newton's
constant
 requires $N\sim P$, i.e., $\b\sim1/2$~\cite{Dimopoulos:2005ac}, in
this paper we therefore focus on the model with $\b=1/2$. By
introducing a convenient dimensionless mass parameter $x_I$ and
corresponding mass spectrum $\tld \ma P(x)$, we show in Fig.
\ref{mass} the mass distribution of $N=1500$ axion fields in the
case of $\b=1/2$. The parameters $x_I$ and $\tld \ma P(x)$ are
defined as
 \bea &&x_I\equiv\f{m_I^2}{a}\;,\qquad \xi\equiv\f{b}{a}\;,\\
 \label{red mp}&&\tld \ma P(x)\equiv a\ma P(m^2)=\f{\sqrt{(\xi-x)(x-1)}}{2\pi\b x\bar x}\;,\qquad
 1<x<\xi\;,\eea
with $\bar x\equiv\s^2/a=1/(1-\sqrt{\b})^2$.

\subsection{Background dynamics}
In the previous subsection we have obtained the mass spectrum for
the $N$ axion fields by virtue of the random matrix theory. Now we
discuss the cosmological background dynamics with the MP spectrum.
In a flat Friedmann-Robertson-Walker universe, the background
dynamics is described by the set of equations
 \bea
 \label{eom1}
 \f{1}{6}\sum_I\B(\dot\vp_I^2+m_I^2\vp_I^2\B )&=&H^2\;,\\
 \label{eom2}
 \f{1}{2}\sum_I\dot\vp_I^2&=&-\dot H\;,\\
 \label{eom3}
 \ddot\vp_I+3H\dot\vp_I+m_I^2\vp_I&=&0\;,
 \eea
where we have taken the units with $M_{pl}=8\pi G=1$. In the slow
roll region
 \be
 \f{\dot\vp_I}{\dot\vp_J}=\f{m_I^2}{m_J^2}\f{\vp_I}{\vp_J}\;,\ee
one has the scaling solution as
 \be\label{sr solution}
 \f{\vp_I(t)}{\vp_I(t_0)}=\B(\f{\vp_J(t)}{\vp_J(t_0)}\B)^{m_I^2/m_J^2}\;,\ee
where $t_0$ denotes the initial time of inflation.
\begin{figure}[h]
  \centering
  \includegraphics[angle=-90,width=9.5cm]{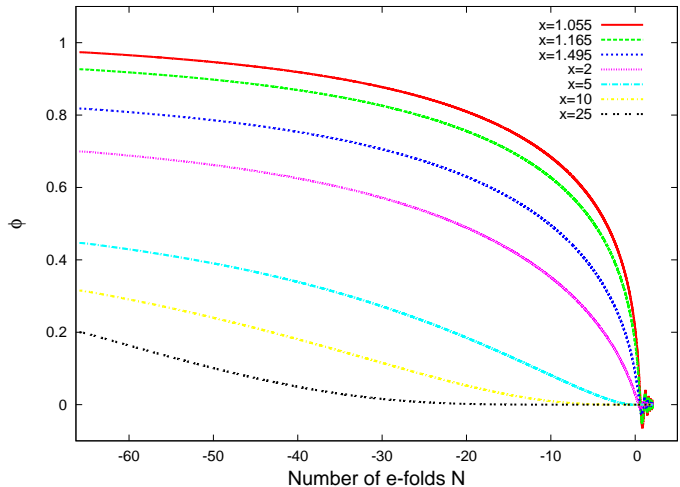}
  \caption{The evolutions of $N=1500$ axion fields versus
          the number of e-folds $\ma{N}$.
          From top to bottom, the evolutions of seven fields $\vp(t;x)$  are plotted with different
          dimensionless mass $x=1.055,1.165,1.495,2,5,10,25$, respectively. }
  \label{evolution}
\end{figure}

In the large $N$ limit, we can deal with the masse distribution of
axions by employing the MP spectrum.  Consequently, the summation
over $I$ in the equations (\ref{eom1}) and (\ref{eom2}) becomes
integrals over mass as
 \bea
 \label{eom4} H^2&=&\f{N}{6}\int_a^b\B[\dot\vp^2(t,m^2)+m^2\vp^2(t,m^2)\B]P(m^2)dm^2\;,\\
 \label{eom5} \dot H&=&-\f{N}{2}\int_a^b\dot\vp^2(t,m^2)P(m^2)dm^2\;.\eea
In principle one can solve the evolution equations (\ref{eom3}),
(\ref{eom4}) and (\ref{eom5}) for one of the $N$ fields, such as the
lightest field, then use the scaling solution (\ref{sr solution}) to
get the solutions for other $N-1$ fields during the slow roll
period.  Because of the complication of the MP spectrum $\ma
P(m^2)$, however, it is very difficult to perform the integration
analytically. In this paper we solve the set of background equations
(\ref{eom1}) and (\ref{eom3}) numerically (see
Fig.~\ref{evolution}), in which we consider $N=1500$ axion fields
evolving from the equal-energy initial configurations
$m_I^2\vp_I^2(t_0)=m_J^2\vp_J^2(t_0)$ with the vacuum expectation
value (vev) of the lightest field, $\vp(t_0,m^2_1)=1$, at the
initial time $t_0$. Our results show that at the initial stage of
inflation, only the heaviest fields (such as $\vp|_{x=25}$) begin
rolling down the potential, after a Hubble time, the heaviest fields
are no longer over-damped. Instead of immediately becoming
under-damped and oscillating they remain critically damped due to
the existence of the lighter fields, and the potential energy of the
heavier fields is dissipated away before it is converted into
kinetic energy. As a result, inflation is mainly sustained by the
lighter fields at the late time and ends till the lightest field is
no longer over-damped.

\section{\label{linear}Linear perturbations}

In this section using the $\d\ma{N}$ formalism, we investigate the
entropy perturbations during inflation at the linear perturbation
level. In the linear cosmological perturbation theory the scalar
perturbations of spacetime are usually parameterized as
 \be\label{metric}
 ds^2=-(1+2\p)dt^2+2a\6_iBdtdx^i+a^2\B[(1-2\psi)\d_{ij}+2\6_i\6_jE\B]dx^idx^j\;,\ee
 where $a$ is the scale factor, $\p$, $\psi$, $B$ and $E$ are four
 scalar perturbations.
 One can define two important gauge invariant quantities as
 \be\label{gaug inv}\z=-\psi-H\f{\d\r}{\dot\r}\;,\qquad
 Q_I=\d\p_I+\f{\dot\p_I}{H}\psi\;.\ee
In the following subsections we  first review the $\d\ma{N}$
formalism and then analyze the entropy perturbations during
inflation by using the method proposed by Tye, Xu and Zhang in
\cite{Tye:2008ef}, and finally calculate the power spectrum,
spectral index and the tensor-to-scalar ratio numerically.

\subsection{Brief review of the $\d \ma{N}$ formalism}

The primordial curvature perturbation $\z(t,\bx)$ on large scales
can be usually calculated by use of the $\d\ma{N}$ formalism
\cite{dN}. (For a multi-field inflation model the $\d\ma{N}$
formalism is nicely reviewed in \cite{Tye:2008ef}.) One of the
essential assumptions of the $\d\ma{N}$ formalism is the
``separate universe assumption'' \cite{Wands:2000dp}, in which
separate Hubble volume evolves like separate
Friedmann-Robertson-Walker universe where density and pressure may
take different values, but are locally homogeneous. Due to the
different e-folding numbers between separate Hubble patches, the
large scale curvature perturbation $\z(t,\bx)$ on the uniform
energy density slice can be expressed as the e-folding number
difference between the uniform energy density slice and the
unperturbed spatially flat slice at the end of inflation
 \be\label{dN1}
 \z(t_E,\bx)=\d\ma{N}=\ma N_e(\vp_I(t_{\ast},\bx),t_E)-\ma N_e^F\;,\ee
where $\ma N_e(\vp_I(t_{\ast},\bx),t_E)$ and $\ma N_e^F$ are the
number of e-folds on the uniform energy density slice and
spatially flat slice respectively. $\vp_I(t_{\ast},\bx)$ denotes
the field configurations at the time of horizon crossing
$t_{\ast}$ and $t_E$ stands for the time at the end of inflation.
In general, $\d\ma{N}$ can be expanded, up to the second order
perturbations, as
 \be\label{dN expandsion}
 \d\ma{N}=N_IQ^I+\half N_{IJ}\B(Q^IQ^J-\la Q^IQ^J\ra\B)+\cdots\;,\ee
where the expansion coefficients are defined as $N_I\equiv \6
N/\6\vp^I$, $N_{IJ}\equiv \6^2 N/\6\vp^I\6\vp^J$ and $Q^I$ is the
perturbation of $\vp^I$ (\ref{gaug inv}) in the spatially flat
gauge.

In the multi-field scenario, it turns out convenient to identify
the effective inflaton field $\s$ as the path length of the
trajectory in the $N$ dimensional field space
 \be\label{sigma}
 \s(t)\equiv\int_{t_{\ast}}^t\sum_{I=1}^N\dot\vp_I\ve_I^{\s}dt\;,\ee
where the vector $\ve_I^{\s}$ is defined by
 \be\label{esigma}
 \ve_I^{\s}\equiv\f{\dot\vp_I}{\dot\s}\;,\qquad\dot\s^2\equiv\sum_I^N\dot\vp_I^2\;.\ee
Furthermore, one introduces other $N-1$ entropy basis vectors
$\ve_s$ to form a set of orthogonal basis $\{\ve_n\}$, where
$(n=\s,s)$ and $s$ denotes the $N-1$ entropy fields in shorthand.
Then the $N$ evolution equations for the background fields
(\ref{eom3}) can be written as the evolution equation for the
effective single field $\s$
 \be\label{eff eom}
 \ddot\s+3H\dot\s+V_{,\s}=0\;,\ee
where the potential gradient in the direction $\ve_I^{\s}$ is
 \be\label{eff potential}
 V_{,\s}\equiv\f{\6V}{\6\s}=\sum_I\ve_I^{\s}\f{\6V}{\6\vp_I}\;.\ee
Thus the unperturbed e-folding number in the $\ve_I^{\s}$
direction can be expressed as\footnote{One can prove that the
e-folding number $\ma{N}$ in
 ($\ref{efold num}$) is equivalent to the usual definition $ \ma{N}=-1/\Mp^2\sum_I\int_{\p_I^{\ast}}^{\p_I^E}V_I/V_{,I} d\p_I$
 as long as the potential takes the decoupled form $V=\sum_IV_I$ and all fields roll monotonically during inflation.
 We thank Jiajun Xu for useful correspondence about this point.}
 \be\label{efold num}
 \ma{N}=\int_{\s_{\ast}}^{\s_E}\f{H}{\dot\s}d\s\;,
 \ee
and at the linear order,  $\d\ma{N}$ reads
 \bea\label{dN expansion2}
 \d\ma{N}&=&- \left.\f{H}{\dot\s} \right|_{t_{\ast}}(\6_I\s^{\ast})Q^I+\le.\f{H}{\dot\s}\ri |_{t_E}(\6_I\s_E)Q^I\nn\\
 &&+\int_{\s_{\ast}}^{\s_E}\f{H}{\dot\s}d\B(\f{\d\s}{\d\vp^I}Q^I\B)-\int_{\s_{\ast}}^{\s_E}
 \f{H}{\dot\s^2}Q^s\6_s\dot\s d\s\;,\eea
where $Q^s$ denotes the field perturbations in entropy directions
$\ve_s$. As pointed out in \cite{Tye:2008ef}, the first term above
comes from the shift in the initial value $\s_{\ast}$, and it
corresponds to the adiabatic perturbation in the single field
case. The second term arises when the uniform energy slice at the
end of inflation is not orthogonal to the background trajectory.
And the third and fourth terms are both dependent on the complete
inflation trajectory after $t_{\ast}$, which reflects the fact
that under the entropy perturbations, the inflaton follows a new
trajectory with different length (the third term) and also
different speed (the fourth term).

For simplicity, in this paper we ignore the contributions from the
second term and rewrite the last two terms using some geometric
tricks{\footnote{The detailed derivations can be found in the
Appendix A of \cite{Tye:2008ef}.}}, then ($\ref{dN expansion2}$)
becomes
 \be\label{dN2}
 \d\ma{N}=-\le.\f{H}{\dot\s}\ri |_{t_{\ast}}Q^{\s}-\int_{t_{\ast}}^{t_E}\f{2H}{\dot\s}\dot\ve_{\s}^IQ_Idt\;,\ee
with
 \be\label{dot e}
 \dot\ve_{\s}^I=-\f{V_{,I}}{\dot\s}+\sum_J\f{V_{,J}}{\dot\s}\ve_{\s}^J\ve_{\s}^I\;.\ee
 One can see from (\ref{dN2}) that, although there exist $N-1$
entropy modes, can only the one which is along the $\dot\ve_{\s}$
direction  seed the curvature perturbation. Therefore we can use
the two-field formalism \cite{twofield} to discuss the entropy
perturbation.

\subsection{Observational predictions}

Now we  calculate the observational predictions of N-flation, such
as the scalar power spectra $P_{\z}$, spectral index $n_s$ and the
tensor-to-scalar ratio $r$. In order to calculate the power
spectrum of the curvature perturbation, it is convenient to move
to the momentum space. The Fourier mode of $\z_{\bk}(t)$ reads
 \bea\label{zetak}
 \z_{\bk}(t)=-\le.\f{H}{\dot\s}\ri |_{t_{\ast}}Q^{\s}(t_{\ast},\bk)-\sum_{I,s}\int_{t_{\ast}}^t
 \f{2H(t')}{\dot\s(t')}\dot\ve_{\s}^I(t')\ve_s^I(t_{\ast})dt'Q^s(t_{\ast},\bk)\;.\eea
Using the $\d\ma{N}$ formalism and choosing the standard
Bunch-Davies vacuum,
 \be\label{vacua}
 \la Q_{\s}Q_{\s}\ra=\f{H^2}{2k^3}\;,\qquad \la Q_sQ_{s'}\ra=\d_{ss'}\f{H^2}{2k^3}\;,\qquad
 \la Q_{\s}Q_s\ra=0\;,\ee
the two-point correlation functions of the curvature perturbation
can be expressed as
 \bea\label{pow}
 \la\z_{\bk_1}(t)\z_{\bk_2}(t)\ra&=&N_{\s}N_{\s}\la Q^{\s}Q^{\s}\ra
 +\sum_{s,s'}N_sN_{s'}\la Q^sQ^{s'}\ra\;,\nn\\
 &=&\le.\f{H^2}{4k^3\e}\ri |_{t_{\ast}}
 +\B(\f{H^2_{\ast}}{2k^3}\B)\sum_{I,J,s}\int_{t_{\ast}}^tdt_1\int_{t_{\ast}}^tdt_2\f{2H(t_1)}{\dot\s(t_1)}\f{2H(t_2)}{\dot\s(t_2)}
 \dot\ve_{\s}^I(t_1)\dot\ve_{\s}^J(t_2)\ve_s^I(t_{\ast})\ve_s^J(t_{\ast})\;,\eea
where the quantities with subscript $\ast$ denote the quantities
take the values at horizon crossing and the slow roll parameters
are defined by
 \be\label{sr para}
 \e\equiv-\f{\dot H}{H^2}\;,\qquad\h\equiv\f{\dot\e}{\e H}\;.\ee
With the help of the orthogonal relation among the entropy basis
vectors
 \be\label{del perpen}
 \sum_s\ve_s^I(t_{\ast})\ve_s^J(t_{\ast})=\d_{\perp}^{IJ}=\d^{IJ}-\ve_{\s}^I(t_{\ast})\ve_{\s}^J(t_{\ast})\;,\ee
the second term of the right hand side of (\ref{pow}) can be
expressed as
 \be\label{pow2}
 \sum_{s,s'}N_sN_{s'}\la
 Q^sQ^{s'}\ra=\B(\f{H^2_{\ast}}{4k^3\e}\B)2\e\ma{N}_{ss}\;,\ee
 with
 \be\label{Nss}
 \ma{N}_{ss}(t,t_{\ast})\equiv\le\{\sum_I\le[\int_{t_{\ast}}^tdt'\f{2H(t')}{\dot\s(t')}\dot\ve_{\s}^I(t')\ri]^2
 -\le[\int_{t_{\ast}}^tdt'\f{2H(t')}{\dot\s(t')}
 \B(\sum_I\dot\ve_{\s}^I(t')\ve_{\s}^I(t_{\ast})\B)\ri]^2\ri\}\;.\ee

 \begin{figure}[h]
  \centering
  \includegraphics[angle=-90,width=9.5cm]{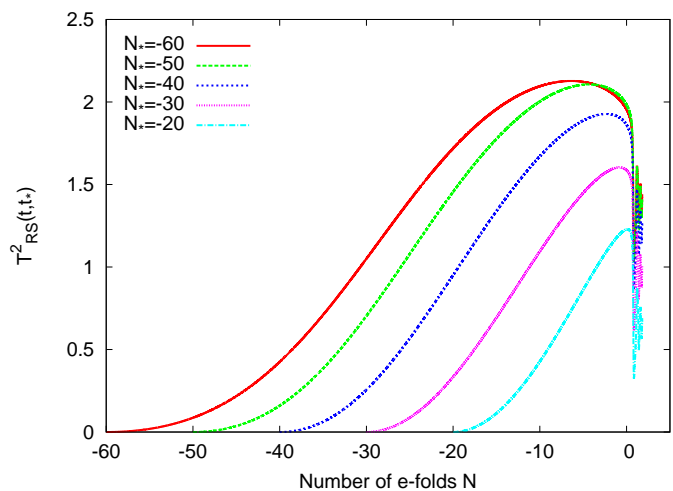}
  \caption{The evolutions of the transfer coefficient $T^2_{\ma{R}\ma{S}}(t,t_{\ast})$ versus the number of
  e-folds ${\cal N}$. From top to bottom, the time evolutions of
  $T^2_{\ma{R}\ma{S}}(t,t_{\ast})$ are plotted for five different wavenumbers, which cross horizon at
  the number of e-folds $\ma{N}_{\ast}=-60,-50,-40,-30,-20,$ respectively. Our results show that the amplitudes of
  spectra decrease with the increasing of the perturbation wavenumber $k$,
  which indicates a red tilt spectrum.}
  \label{trs}
 \end{figure}

For a two-field model one can argue on a very general ground that
the time dependence of curvature and entropy perturbations in the
large-scale limit can always be described by \cite{twofield}
 \be\label{RSevol}
 \dot\ma{R}=\a H\ma{S}\;,\qquad \dot\ma{S}=\b H\ma{S}\;,
 \ee
where~\footnote{In this paper, $\ma{R}$ is denoted as $\z$ and the
sound speed $c_s=1$ because the
 kinetic term is canonical in the N-flation model.}
 \be\label{adi entr}
 \ma{R}\equiv\f{H}{\dot\s}Q_{\s}\;,\qquad \ma{S}\equiv c_s\f{H}{\dot{\s}}Q_s\;,
 \ee
and $\a$, $\b$ are two time dependent dimensionless functions.
Integrating (\ref{RSevol}), one can obtain a general form of the
transfer matrix which relates the curvature and entropy
perturbations generated at horizon crossing $t_{\ast}$ to those at
some later time $t$
 \be
        \lrp{\begin{array}{c}
               \mathcal{R} \\
               \mathcal{S}
             \end{array}
         } = \lrp{ \begin{array}{cc}
                     1 & T_{\mathcal{R}\mathcal{S}} \\
                     0 & T_{\mathcal{S}\mathcal{S}}
                   \end{array}
          } \lrp{\begin{array}{c}
               \mathcal{R} \\
               \mathcal{S}
             \end{array}
         }_{\ast} \,,\label{transf}
 \ee
where
 \be
        T_{\mathcal{S}\mathcal{S}}(t,t_{\ast}) = \exp\left\{{\int_{t_{\ast}}^t dt'\, \beta(t')H(t')
        }\right\}\,,\qquad T_{\mathcal{R}\mathcal{S}}(t,t_{\ast}) = \int_{t_{\ast}}^t
        dt'\, \alpha(t') T_{\mathcal{S}\mathcal{S}}(t',t_{\ast})
        H(t') \;.
 \ee
As we have shown in (\ref{dN2}), for the multi-field model, can
only one entropy mode, which is along the $\dot\ve_{\s}$
direction, contribute to the overall curvature perturbation if we
neglect the torsion in the background trajectory. That is to say,
we can take $\ma{S}$ as the entropy perturbation in the
$\dot\ve_{\s}$ direction, namely, at the linear perturbation level
the multi-field model is effectively equivalent to the two-field
model. The relation ($\ref{transf}$) is still valid in the
N-flation model.

With the above results, the power spectrum of the primordial
curvature perturbation can be expressed as
 \be\label{pow3}
 P_{\z}(t,t_{\ast},\bk)\equiv\f{k^3}{2\pi^2}\la\z_{\bk}\z_{\bk'}\ra
 =\f{H^2_{\ast}}{8\pi^2\e_{\ast}}\B[1+T_{\ma{R}\ma{S}}^2(t,t_{\ast})\B]\;,\ee
where the transfer coefficient
 \be\label{TRS}
 T_{\ma{R}\ma{S}}^2(t,t_{\ast})=2\e_{\ast}\ma{N}_{ss}(t,t_{\ast})\;,\ee
measures the contribution to the overall curvature perturbation
from the entropy modes. The first constant term in (\ref{pow3})
comes from the pure adiabatic perturbation at the horizon
crossing, and the second term, which is time dependent, describes
the entropy contributions to the curvature perturbation. Because
of the existence of the entropy modes, the curvature perturbation
does not conserve after the horizon crossing, and the second term
does characterize the time evolution of the spectrum from the
horizon crossing to the end of inflation. In Fig.~\ref{trs}, we
show the time evolutions of the power spectra with different
wavelengthes numerically.  The results show that the amplitudes of
spectra decrease with the increasing of the perturbation
wavenumber $k$, which implies a red tilt spectrum.

In order to further confirm the above analysis about the power
spectrum, we
 calculate the spectral index explicitly.  The spectral index
 turns out to be
 \bea\label{index}
 n_S-1&\equiv&\f{d\log P_{\z}(t_E,t_{\ast},k)}{d\log k}=\f{d\log H^2_{\ast}}{d\log k}-\f{d\log \e_{\ast}}{d\log
 k}+\f{d\log \B[1+2\e_{\ast}\ma{N}_{ss}(t_E,t_{\ast})\B]}{d\log k}\nn\\
 &=&-2\e-\h+\f{2\h\e\ma{N}_{ss}+2\e\dot{\ma{N}}_{ss}/H}{\B(1+2\e\ma{N}_{ss}\B)}\;,\eea
where we have fixed $t=t_E$ and the $\dot\ma{N}_{ss}$ reads
 \bea\label{Nssdot}
 \dot{\ma{N}}_{ss}(t_E,t_{\ast})&=&\f{d\ma{N}_{ss}}{dt_{\ast}}=\sum_I\f{-4H_{\ast}}{\dot\s_{\ast}}
 \dot{\ve}^I_{\s}(t_{\ast})\B[\int_{t_{\ast}}^{t_E}
 dt'\f{2H(t')}{\dot{\s}(t')}\dot{\ve}^I_{\s}(t')\B]\nn\\
 &&-2\B[\int_{t_{\ast}}^{t_E}dt'\f{2H(t')}{\dot\s(t')}\B(\sum_I\dot\ve^I_{\s}(t')\ve^I_{\s}(t_{\ast})\B)\B]
 \cdot\B[\int_{t_{\ast}}^{t_E}dt'\f{2H(t')}{\dot\s(t')}\B(\sum_J\dot\ve^J_{\s}(t')\dot\ve^J_{\s}(t_{\ast})\B)\B]\;.
 \eea

 \begin{figure}[h]
 \centering
  \includegraphics[angle=-90,width=9.5cm]{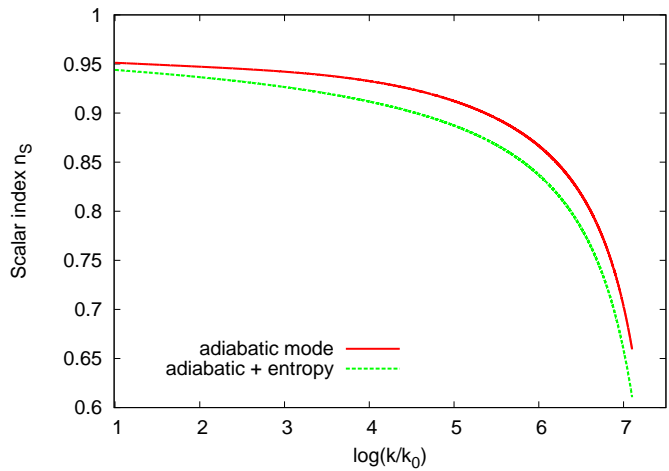}
  \caption{The scalar index $n_S$ versus the logarithm of dimensionless comoving wavenumber $\log
  k/k_0$, where $k_0$ stands for the comoving wavenumber of the mode which crosses horizon at the number of
  e-folds  $\ma{N}_{\ast}=-60$.
  The solid (red) curve denotes the contributions to $n_S$ from
  the pure adiabatic component, while the dashed (green) one includes both
adiabatic and entropy components. }
  \label{index}
 \end{figure}

 We plot in Fig.~\ref{index} the scalar index $n_S$ versus the
logarithm of dimensionless comoving wavenumber $\log  k/k_0$. It
can be seen from the figure  that compared to the pure adiabatic
case, the index becomes smaller after including the entropy
components. This numerical results are in agreement with the
analytic ones in~\cite{Battefeld:2006sz,Choi:2008et}.

\begin{figure}[h]
 \centering
  \includegraphics[angle=-90,width=9.5cm]{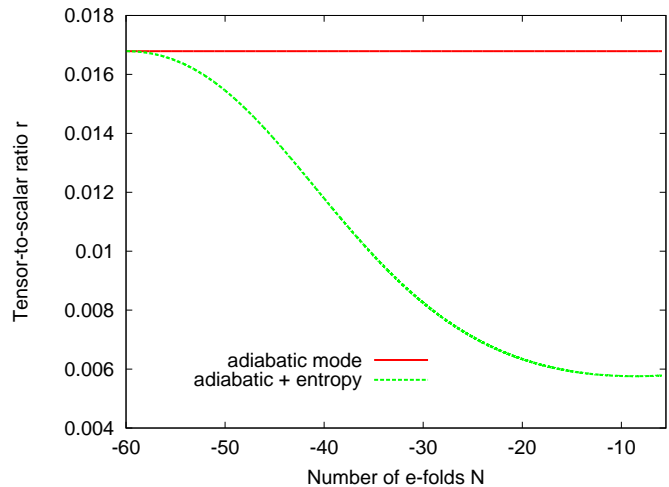}
  \caption{The tensor-to-scalar ratio $r$ for the $k_0$ mode which crosses horizon at the e-folding number
  $\ma{N}_{\ast}=-60$. The solid (red) curve denotes
  the ratio $r=\e_{\ast}$ calculated from the pure adiabatic perturbation,
  while the dashed (green) one
  $r=\e_{\ast}\sin^2\Theta(t,t_{\ast})$ describes the evolution of the ratio from horizon crossing to the
  end of inflation. }
  \label{tsratio}
 \end{figure}

Finally we  discuss the modified consistency relation
\cite{Wands:2007bd,Bartolo:2001rt} in N-flation. Because the
tensor perturbation is decoupled from the scalar one at the linear
order, the gravitational wave power spectrum is frozen-in on large
scales as what happens in the single-field model
 \be\label{gw}
 P_T=P_T|_{\ast}=\f{8H^2_{\ast}}{4\pi^2}\;.\ee
We define the tensor-to-scalar ratio for a given $k$ mode which
crosses horizon at e-folding number $\ma{N}_{\ast}$ as
 \bea\label{ratio}
 r(t,t_{\ast})&=&\f{P_T(t_{\ast})}{16P_{\z}(t,t_{\ast})}
 =\f{\e_{\ast}}{1+T^2_{\ma{R}\ma{S}}(t,t_{\ast})}\nn\\
 &=&\e_{\ast}\sin^2\Theta(t,t_{\ast})\;,\qquad\qquad (t_{\ast}\leq t\leq t_E)\;,\eea
where we have introduced a dimensionless correlation angle
$\Theta$
 \be\label{angle}
 \sin\Theta=\f{1}{\sqrt{1+T^2_{\ma{R}\ma{S}}}}\;.\ee
We can see from (\ref{ratio}) that, after taking into account the
entropy perturbations, the tensor-to-scalar ratio
($r=\e\sin^2\Theta$) is always smaller than the one for the case
of the pure adiabatic perturbation ($r=\e_{\ast}$). In Fig.
\ref{tsratio} we  show the tensor-to-scalar ratio of the $k_0$
mode which crosses horizon at e-folding number
$\ma{N}_{\ast}=-60$. The results show that the ratio
($r\simeq0.006$) is greatly suppressed by the entropy modes at the
end of inflation.

\section{\label{concl}Conclusion}

In this paper we studied numerically the dynamics of N-flation. At
the background evolution level we investigated the evolution of
$N$ axions with the Marc\v enko-Pastur mass distribution, and we
found that at the initial stage of inflation, only the heaviest
fields begin sliding down the potential, after a Hubble time the
heaviest fields are no longer over-damped. Instead of immediately
becoming under-damped and oscillating they remain critically
damped due to the existence of the lighter fields, and all the
potential energy of the heavier fields is dissipated away before
it is converted into the kinetic energy. As a result the inflation
is mainly sustained by the lighter fields at the late time and
ends until the lightest field were no longer over-damped.

At the linear perturbation level, we calculated the corrections of
entropy perturbations to the power spectrum of the overall
curvature perturbation $P_{\z}$ by use of the $\d\ma{N}$
formalism. We obtained an analytic form of the transfer
coefficient $T^2_{\ma{R}\ma{S}}$, which describes the correlation
between the curvature and entropy perturbations, and investigated
its behavior numerically. Our results show that the entropy
perturbations cannot be neglected in the N-flation model, because
the amplitude of entropy components is approximately in the same
order as the adiabatic one at the end of inflation
$T^2_{\ma{R}\ma{S}}\sim\ma{O}(1)$. Then we calculated the spectral
index $n_S$ and found that the index becomes smaller once the
entropy modes are included, i.e., the spectrum becomes redder than
the pure adiabatic one. Finally we studied the modified
consistency relation for the N-flation model and found that the
tensor-to-scalar ratio ($r\simeq0.006$) is greatly suppressed by
the entropy modes, compared to the pure adiabatic one
($r\simeq0.017$) at the end of inflation.

In this paper we only considered the entropy perturbations from
the third and fourth terms in the right hand side of (\ref{dN
expansion2}), which depend on the whole background trajectory in
field space, while ignored the corrections to the curvature
perturbation which would arise when the uniform energy slice at
the end of inflation was not orthogonal to the background
trajectory. In addition, the additional power in the curvature
perturbation, which may be produced by the (p)reheating process,
is also out of the discussion in this paper.

\begin{acknowledgments}
BH thanks Seoktae Koh for useful discussions and Jiajun Xu for
helpful correspondence. BH and RGC are supported in part by the
National Natural Science Foundation of China under Grant Nos.
10535060, 10821504 and 10975168, and by the Ministry of Science
and Technology of China under Grant No. 2010CB833004. YSP is
supported in part by NSFC under Grant No: 10775180 and by the
Scientific Research Fund of GUCAS,
\end{acknowledgments}

\appendix

\end{document}